\documentclass[%
preprint,
superscriptaddress,
 amsmath,amssymb,
 aps,
 prl,
]{revtex4-2}

\usepackage[dvipdfmx]{graphicx}
\usepackage{dcolumn}
\usepackage{bm}

\usepackage{color}
\usepackage{ulem}
\newcommand{\add}[1]{\textcolor{black}{#1}}
\newcommand{\del}[1]{}

\begin{document}


\title{Robust topological surface states in skyrmion-host magnets $\mathrm{Eu(Ga,Al)_4}$: evidence for dual topology}

\author{Yuki Arai}
\affiliation{Department of Physics, Tohoku University, Sendai 980-8578, Japan}

\author{Kosuke Nakayama}
\thanks{Corresponding authors:\\
k.nakayama@arpes.phys.tohoku.ac.jp\\
segawa@cc.kyoto-su.ac.jp\\
t-sato@arpes.phys.tohoku.ac.jp}
\affiliation{Department of Physics, Tohoku University, Sendai 980-8578, Japan}

\author{Takemi Kato}
\affiliation{Advanced Institute for Materials Research (WPI-AIMR), Tohoku University, Sendai 980-8577, Japan}
\affiliation{Quantum Materials Science Unit, Okinawa Institute of Science and Technology (OIST), Okinawa 904-0495, Japan}

\author{Tomonori Nakamura}
\affiliation{Quantum Materials Science Unit, Okinawa Institute of Science and Technology (OIST), Okinawa 904-0495, Japan}

\author{Asuka Honma}
\affiliation{Department of Physics, Tohoku University, Sendai 980-8578, Japan}

\author{Seigo Souma}
\affiliation{Advanced Institute for Materials Research (WPI-AIMR), Tohoku University, Sendai 980-8577, Japan}
\affiliation{Center for Science and Innovation in Spintronics (CSIS), Tohoku University, Sendai 980-8577, Japan}

\author{Kenichi Ozawa}
\affiliation{Institute of Materials Structure Science, High Energy Accelerator Research Organization (KEK), Tsukuba, Ibaraki 305-0801, Japan.}

\author{Kiyohisa Tanaka}
\affiliation{UVSOR Synchrotron Facility, Institute for Molecular Science, Okazaki 444-8585, Japan}
\affiliation{School of Physical Sciences, The Graduate University for Advanced Studies (SOKENDAI), Okazaki 444- 8585, Japan}

\author{Daisuke Shiga}
\affiliation{Institute of Multidisciplinary Research for Advanced Materials (IMRAM), Tohoku University, Sendai 980-8577, Japan}

\author{Hiroshi Kumigashira}
\affiliation{Institute of Multidisciplinary Research for Advanced Materials (IMRAM), Tohoku University, Sendai 980-8577, Japan}

\author{Yoshinori Okada}
\affiliation{Quantum Materials Science Unit, Okinawa Institute of Science and Technology (OIST), Okinawa 904-0495, Japan}

\author{Kouji Segawa}
\thanks{Corresponding authors:\\
k.nakayama@arpes.phys.tohoku.ac.jp\\
segawa@cc.kyoto-su.ac.jp\\
t-sato@arpes.phys.tohoku.ac.jp}
\affiliation{Department of Physics, Kyoto Sangyo University, Kyoto 603-8555, Japan.}

\author{Takafumi Sato}
\thanks{Corresponding authors:\\
k.nakayama@arpes.phys.tohoku.ac.jp\\
segawa@cc.kyoto-su.ac.jp\\
t-sato@arpes.phys.tohoku.ac.jp}
\affiliation{Department of Physics, Tohoku University, Sendai 980-8578, Japan}
\affiliation{Advanced Institute for Materials Research (WPI-AIMR), Tohoku University, Sendai 980-8577, Japan}
\affiliation{Center for Science and Innovation in Spintronics (CSIS), Tohoku University, Sendai 980-8577, Japan}
\affiliation{International Center for Synchrotron Radiation Innovation Smart (SRIS), Tohoku University, Sendai 980-8577, Japan}
\affiliation{Mathematical Science Center for Co-creative Society (MathCCS), Tohoku University, Sendai 980-8578, Japan}

\date{\today}

\begin{abstract}
The interplay between real-space topology such as magnetic skyrmions and momentum-space topology characterized by topological surface states (TSSs) is predicted to realize novel phenomena and functionalities, yet materials hosting both topologies are scarce. Skyrmion-hosting helimagnet family $\mathrm{EuGa_2Al_2}$ and $\mathrm{EuAl_4}$ has been a prime candidate for such a dual-topology system, but conclusive evidence for its momentum-space topology has remained elusive. We provide this evidence by directly observing TSSs that stem from bulk Dirac nodal lines using high-resolution angle-resolved photoemission spectroscopy. These TSSs are exceptionally robust against various perturbations such as a 2$\times$1 surface reconstruction, a chemical change in the termination of the crystal surface, and the onset of helical antiferromagnetic order. Crucially, below the N\'{e}el temperature, we observe replica bands driven by the magnetic ordering. Moreover, we demonstrate clear surface-termination dependence of this magneto-topological coupling. Our findings establish $\mathrm{Eu(Ga_{1-\mathnormal{x}}Al_{\mathnormal{x}})_4}$ as a dual-topology material and offer a rare platform to explore and control the interaction between the two fundamental topological realms.
\end{abstract}

\maketitle

Nontrivial topology plays a pivotal role in modern condensed matter physics, stabilizing exotic phases of matter and emergent phenomena. Magnetic skyrmions, particle-like spin vortices, are a primary example of topological objects in real space \cite{RoblerNat2006, MuhlbauerScience2009, NagaosaNatTech2013}. Their nontrivial spin texture gives rise to a large emergent electromagnetic field, leading to fascinating phenomena such as the topological Hall and Nernst effects \cite{LeePRL2009, NeubauerPRL2009, ShiomiPRB2009}. Their topological protection ensures remarkable robustness against perturbations, offering great potential for low-power spintronic applications \cite{FertNatTech2013, SampaioNaTech2013, Gobel2021Beyond}. In parallel, topological phases characterized by momentum-space topology have revolutionalized our understanding of quantum materials. In topological insulators \cite{KanePRL1, KanePRL2, BernevigPRL2006}, for instance, nontrivial band topology originating from band inversion leads to gapless topological surface states (TSSs) \cite{XiaNP2009, ChenScience2009}, underpinning unique quantum transport phenomena \cite{HasanRMP2010, QiRMP2011, AndoJPSJ2013}. The concept of momentum-space topology has since expanded to topological semimetals and superconductors \cite{HasanRMP2010, QiRMP2011, AndoJPSJ2013, ArmitageRMP2018, Sato2017Review}, broadening the landscape of topological quantum matter.

While these two notions of topology, real-space and momentum-space, have each spurred extensive research, materials where they coexist and intertwine remain extremely rare. Establishing such ``dual-topology" systems is of fundamental importance, as their interplay is predicted to host fascinating phenomena such as charging of skyrmions by dressing them with TSSs, emergent gauge coupling between spin and charge textures, and efficient and novel topological spin control \cite{HurstPRB2015, AndrikopoulosJAP2016, KurebayashiPRR, MoyaPRB2023, WuAdvMat2020}. Yet a candidate material platform where both topological orders are unambiguously verified by experiments has not been firmly established.

In this context, the $\text{Eu}(\text{Ga}_{1-x}\text{Al}_x)_4$ family [crystal structure and corresponding Brillouin zone are shown in Figs. 1(a) and 1(b), respectively] has recently emerged as a promising candidate for realizing such dual topology \cite{NakamuraJPSJ2015, TakagiNC2022, LeiNC2023, MoyaPRB2023, ShangPRB2021, ShangReview2024}. Previous studies have revealed that $\text{EuAl}_4$ ($x = 1.0$) hosts multiple nanometer-scale skyrmion phases under applied magnetic fields \cite{TakagiNC2022, ShangPRB2021}, and the topological Hall effect observed across a wide composition range ($0.5 \le x \le 1$) attests to its nontrivial real-space topology \cite{ShangPRB2021, MoyaPRB2023}. At the same time, the $\text{Eu}(\text{Ga}_{1-x}\text{Al}_x)_4$ series exhibits remarkable momentum-space features linked to its electronic structure. The Ruderman-Kittel-Kasuya-Yosida interaction, originating from Fermi-surface ($\text{FS}$) nesting, has been identified as the driving force of multi-$Q$ helical magnetic orders \cite{AraiNC2026} that produce skyrmion lattices \cite{TakagiNC2022, HayamiJPSJ2022, HayamiJPM2023}. This implies a direct coupling between fermiology and magnetic topology. Indeed, enormous unsaturated magnetoresistance up to $200,000\%$ has been discovered in $\text{EuGa}_4$ ($x = 0$) and attributed to the formation of a topological Weyl nodal line near the Fermi level ($E_{\rm F}$) \cite{LeiNC2023}. Given the same crystal symmetry, $\text{Eu}(\text{Ga}_{1-x}\text{Al}_x)_4$ at $0.5 \le x \le 1.0$, where skyrmion formation and topological Hall effect have been reported, are expected to host a similar nodal line [i.e., Dirac nodal line (DNL) in the absence of magnetic field]. However, the direct experimental evidence of momentum-space topology, namely the corresponding $\text{TSSs}$, has been absent, despite intensive experimental studies on the electronic structure \cite{LeiNC2023, AraiNC2026, KobataJPSJ2016, EatonPRB2024, MiaoPRX2024, LiArXiv2025}. This missing piece prevents the definitive identification of $\text{Eu}(\text{Ga}_{1-x}\text{Al}_x)_4$ as a true dual-topology system and the fundamental interplay between its two topological realms has remained unexplored.

In this Letter, using high-resolution angle-resolved photoemission spectroscopy (ARPES), we report the direct observation of the $\text{TSSs}$ associated with the bulk $\text{Dirac}$ nodal lines in $\text{EuGa}_2\text{Al}_2$ ($x = 0.5$; the N\'{e}el temperature $T_{\rm N}$ = 19.0 K) and $\text{EuAl}_4$ ($x = 1.0$; $T_{\rm N}$ = 15.4 K). We also uncover their exceptional robustness against surface and magnetic perturbations and reveal a direct coupling between the helical magnetism and the electronic structure, including the TSS, thereby completing the picture of $\text{Eu}(\text{Ga}_{1-x}\text{Al}_x)_4$ as a dual-topology magnet.

We begin with the investigation of the electronic states in the paramagnetic state of $\text{EuGa}_2\text{Al}_2$ using surface-sensitive $\text{ARPES}$ measurements with vacuum ultraviolet ($\text{VUV}$) photons (see Section 1 of Supplemental Material for details on the experimental conditions \add{and Section 2 of Supplemental Material for a Fast-Fourier-transformation filtering method applied to ARPES data to remove mesh grid artifacts} \cite{SM}). Figure 1(c) shows the $\text{FS}$ mapping at $T$ = 30 K with $h\nu = 113\,\text{eV}$ on a surface terminated by the $\text{GaAl}$ tetrahedral layer with the Ga-Al-Ga ordered stacking arrangement \cite{StavinohaPRB2018} (surface termination will be demonstrated later in Fig. \add{3}). The high surface sensitivity is confirmed by three key characteristics. First, $\text{FSs}$ in the first and second $\text{Brillouin}$ zones ($\text{BZs}$) appear similar [e.g., a large star-shaped feature (labeled $\text{S1}$; yellow solid line) is commonly seen], despite a $\pi$-phase shift in $k_z$ between the neighboring bulk $\text{BZs}$ [see Fig. 1(b)]. This reflects strong $k_z$ broadening that originates from the short mean free path of photoelectrons excited by $\text{VUV}$ photons, i.e., the emission of photoelectrons exclusively from the surface region \add{[although some $k_z$ selectivity remains, as seen later in Figs. 1(f) and 1(g)]}. Second, the $\text{FSs}$ exhibit $\text{C}_2$ symmetry, rather than the $\text{C}_4$ symmetry of the bulk tetragonal lattice [see the calculated bulk $\text{C}_4$-symmetric $\text{FSs}$ at $k_z = \pi$ in Fig. 1(d)], due to the appearance of replica $\text{FSs}$ shifted by $\pm\pi/a$ along the $k_y$ direction [e.g., see a replica of the $\text{S1}$ $\text{FS}$ ($\text{S1}'$) traced by yellow dashed lines]. This replica is associated with a $2\times1$ surface reconstruction, as supported by our scanning tunneling microscopy ($\text{STM}$) results [Fig. 1(e)]. Third, and most importantly, we observe $\text{FSs}$ that cannot be ascribed to either the bulk bands or their $2\times1$ replica bands. For example, while the bulk band calculations predict only two $\text{FS}$ sheets whose vertices extend toward the $\bar{\rm X}$ point of the surface $\text{BZ}$ [Fig. 1(d); see Section \add{3} of Supplemental Material for details on the calculations \cite{SM}], our experiment clearly resolves three sheets ($\text{S1}$, $\text{B1}$, and $\text{B3}$). This additional feature serves as evidence for the $\text{SS}$ formation.

To investigate the $\text{SSs}$ in more detail, we present in Figs. 1(f) and 1(g) the $\text{ARPES}$-intensity plots measured along the $\bar{\Gamma}$-$\bar{\rm X}$ cut at $h\nu = 113\,\text{eV}$ and $90\,\text{eV}$ that probe $k_z \sim 0$ and $\pi$, respectively \add{(see Section 4 of Supplemental Material for details on the normal-emission measurements \cite{SM})}. \add{Many bands ($\text{S1}$, $\text{S1}'$, $\text{B1}$, $\text{B2}$, $\text{B3}$, $\text{B4}$, $\text{B5}$, and $\text{B6}$) are identified near the Fermi level ($E_{\rm F}$).} The $\text{S1}'$ band appears by folding the $\bar{\rm X}$-centered hole-like dispersion of the $\text{S1}$ band from $k_y = \pm\pi/a$ ($\bar{\rm X}$-$\bar{\rm M}$ cut) onto the $\bar{\Gamma}$ point, which contrasts with the $\bar{\rm X}$-centered electron-like $\text{S1}$-band dispersion at $k_y = 0$ ($\bar{\Gamma}$-$\bar{\rm X}$ cut). In addition, at a higher binding energy ($E_{\rm B}$) of $\sim0.5\,\text{eV}$, there is a relatively flat band ($\text{S2}$) irrespective of $k_z$. A comparison of the experimental band dispersions (lines and circles) with the calculated bulk band projection in Fig. 1(h) reveals that while many bands (indicated by lines) correspond to the bulk states, those labeled $\text{S1}$ and $\text{S2}$ (circles) have no bulk counterparts, and extend into the gapped region of the projection. Our slab-model calculations in Fig. 1(i) confirm their $\text{SS}$ origin. Notably, a near-$E_{\rm F}$ electron-like $\text{SS}$ bottomed at $k_x = \pi/a$ in the calculations is attributed to the experimental $\text{S1}$ band (see Section \add{5} in Supplemental Material for the calculated $\text{FSs}$ \cite{SM}\add{; note that the quantitative energy discrepancy for the $\text{S1}$ band between ARPES and DFT is likely due to differences in the surface potential, as our model does not explicitly incorporate the observed 2$\times$1 surface reconstruction whose atomic details are still unclear}). In addition, a relatively flat $\text{SS}$ at $E_{\rm B} \sim 0.5\,\text{eV}$ corresponds to the $\text{S2}$ band. The absence of discernible $h\nu$ dependence in the $\text{S1}$ and $\text{S2}$ band dispersion is also consistent with their two-dimensional $\text{SS}$ nature [compare Figs. 1(f) and 1(g); see also Section \add{6} in Supplemental Material \cite{SM}]. The key question is the topological character of these $\text{SSs}$. $\text{Eu}(\text{Ga}_{1-x}\text{Al}_x)_4$ is a material family that hosts $\text{DNLs}$ \cite{LeiNC2023}. For example, the $\text{B2}$ and another bulk band cross to form a nodal point at $E_{\rm B} \sim 0.65$ eV and produce a $\text{DNL}$ in $k_x$-$k_y$ space, as marked by a yellow arrow in Fig. 1(h) [see also a red dot Fig. 1(i)]. Crucially, the $\text{S2}$ band is connected to this bulk $\text{DNL}$, and forms a drumhead-like dispersion. This connectivity is an essential characteristic of $\text{TSSs}$ in $\text{Dirac}$ semimetals and strongly supports the nontrivial topological origin \add{(see Section 7 of Supplemental Material for the role of spin-orbit coupling on the band dispersion \cite{SM})}.

\add{To evaluate the topological property, we have performed topological invariant calculations. First, we calculated the Berry phase $\gamma_n$ along a closed path encircling the bulk $\text{DNL}$ of current focus [labeled DNL1 in Fig. 2(a)], to which the S2 surface state is connected. We found that the Berry phase is quantized to $\pi$ for the $\text{DNL1}$, which supports the nontrivial topological origin of the $\text{S2}$ band. Furthermore, we calculated the Zak phase $\theta$, which is a topological invariant often employed for topological semimetals, by integrating the Berry connection along the $k_z$ path for each $k_x$ point, as $\theta(k_x) = -i\sum\limits_{n}\int_{-\pi/c}^{\pi/c}{\left\langle u_n(\mathbf{k})\middle|\partial k_z\middle| u_n(\mathbf{k})\right\rangle d k_z}$, where $u_n(\mathbf{k})$ is the Bloch function of $n$-th band. To elucidate how the surface state emerges from the DNL1, the summation was performed over the bands up to those constituting the DNL1. Our calculations [Fig. 2(b)] reveal that the Zak phase is $\pi$ in the $k$ region outside the $\text{DNL1}$, but not inside [note that the Zak phase shows a jump not only at the $k_x$ point of the DNL1 ($\sim$0.35 $\pi/a$) but also at other $k_x$ ($\sim$0.15 and $\sim$0.85 $\pi/a$) because additional topological $\text{DNLs}$ [$\text{DNL2}$ and $\text{DNL3}$ in Fig. 2(a)] exist between the bands forming the DNL1. This result is consistent with our experimental observation that the $\text{S2}$ band emerges at a large-$k$ region away from the $\text{DNL1}$, corroborating the $\text{TSS}$ nature of the $\text{S2}$ state.}

Our study also establishes the remarkable robustness of these $\text{TSSs}$. As shown in Figs. 1(c) and 1(f), the $\text{TSSs}$ persist despite the presence of the $2\times1$ surface reconstruction. This is a clear manifestation of the topological protection. We have also found that these $\text{SSs}$ persist robustly under more drastic changes to the surface structure, specifically, a change in the surface termination layer, as described in the following.

$\text{EuGa}_2\text{Al}_2$ has a layered crystal structure composed of $\text{GaAl}$ tetrahedral layers and $\text{Eu}$ layers [Fig. \add{3}(a)], so that cleaving is expected to produce the surfaces terminated by either layer. Correspondingly, by scanning a micro-focused photon beam ($\phi \sim$ 10 $\mathrm{\mu}$m) \cite{KitamuraRSI2022} across the cleaved surface and imaging the spatial distribution of $\text{Ga}$ core-level intensities, we identified two distinct domains [red and blue areas in Fig. \add{3}(b)]. Because the red region exhibits weaker $\text{Ga}$-derived intensity (stronger $\text{Eu}$-derived intensity) than the blue region (see Section \add{8} in Supplemental Material \cite{SM}), we identify the red area as $\text{Eu}$-rich ($\text{Eu}$-terminated) domain and the blue area as $\text{Eu}$-poor ($\text{GaAl}$-terminated) domain. The $\text{FS}$ in the paramagnetic phase ($T$ = 30 K) on the $\text{Eu}$-terminated surface [Fig. \add{3}(f)] exhibits several striking differences from that of the $\text{GaAl}$ termination [Fig. \add{3}(c), reproduced from Fig. 1(c)]: it preserves the $1\times1$ structure due to the absence of surface reconstruction (Section \add{9} in Supplemental Material \cite{SM}), and a one-dimensionally-elongated feature ($\text{S3}$) emerges near the $\bar{\rm X}$ point. The band dispersion along the $k_x$ axis on the $\text{Eu}$-terminated surface [Fig. \add{3}(g)] reveals that, around the $\bar{\rm X}$ point, a nearly flat $\text{S3}$ band has appeared near $E_{\rm F}$ (blue line). Our slab calculations for the $\text{Eu}$ termination [Fig. \add{3}(h)] successfully reproduce the $\text{S3}$ band. It is noted that intense signal of the $\text{B2}$ band (blue dashed line) in Fig. \add{3}(g) likely signifies an overlap with the $\text{S3}$ band that disperses to higher $E_{\rm B}$'s away from $\bar{\rm X}$ and connects to a bulk DNL indicated by a red dot in Fig. \add{3}(h). The similar dispersion relation and terminating ($E$, \textbf{k}) point of the S3 band with those of the S2 band indicates its TSS nature. Our observation of $\text{TSSs}$ on two chemically and structurally different surfaces unambiguously demonstrates that their formation is a robust and intrinsic property of the bulk topology. The $\text{SS}$ origin is further supported by bulk-sensitive $\text{soft x-ray}$ $\text{ARPES}$ data [Figs. \add{3}(i)--\add{3}(k)], where all $\text{SS}$-derived features are absent \cite{AraiNC2026}.

The observed $\text{TSSs}$ are also robust against magnetic order. Figures \add{4}(a) and \add{4}(b) show the band dispersion along the $k_x$ axis measured on the $\text{Eu}$-terminated surface above and below $T_{\rm N}$, respectively, demonstrating that the $\text{S3}$ band clearly persists deep into the magnetic-ordered phase. More strikingly, the magnetic order actively couples to the electronic states. Below $T_{\rm N}$, a replica band shifted by $Q$ emerges [compare the constant energy contours at $E_{\rm B} = 0.5\,\text{eV}$ in Figs. \add{4}(c) and \add{4}(d); see also Fig. \add{4}(e) for the band dispersion along cut 1 in Fig. \add{4}(d)], and this $Q$ vector matches the magnetic ordering vector [Fig. \add{4}(f)] \cite{MoyaPRM2022, VibhakarPRB2023}. These observations provide direct evidence of band folding, including that of the TSSs, driven by the magnetic order (see also Supplemental Material Section \add{10} \cite{SM}). This magnetic band folding was not clearly observed on the $\text{GaAl}$-terminated surface, suggesting that the interaction is mediated by the local magnetic moments of the $\text{Eu}$ atoms at the surface.

Finally, to demonstrate the generality of our findings, we investigated the $\text{Al}$-end member, $\text{EuAl}_4$. The $\text{FS}$ and the band dispersion for an $\text{Al}$-terminated surface [Figs. \add{5}(a) and \add{5}(b)] show a unidirectional band folding due to the $2\times1$ surface reconstruction and, importantly, $\text{S1}$ and $\text{S2}$ bands analogous to those in $\text{EuGa}_2\text{Al}_2$ [see also slab calculations in Fig. \add{5}(c)]. This confirms that the nontrivial topology in momentum space is a fundamental characteristic of the $\text{Eu}(\text{Ga}_{1-x}\text{Al}_x)_4$ family for $0.5 \le x \le 1.0$.

Our findings provide the missing experimental evidence that completes the picture of $\text{Eu}(\text{Ga}_{1-x}\text{Al}_x)_4$ as a rare material system exhibiting dual topologies in both real and momentum spaces. The significance of this discovery is highlighted when compared to other $\text{DNL}$ semimetals. For instance, while materials such as the $\text{ZrSiS}$ family are typical examples of $\text{DNL}$ semimetals \cite{SchoopNC2016, TakanePRB2016}, the coexistence of their $\text{SSs}$ with other quantum orders is rare, and the $\text{SSs}$ themselves may be unstable \cite{BoukhvalovAFM2019}. In stark contrast, the $\text{TSSs}$ in $\text{Eu}(\text{Ga}_{1-x}\text{Al}_x)_4$ not only coexist with magnetic order but also exhibit remarkable robustness. The distinction extends to magnetic $\text{DNL}$ systems hosting $\text{TSSs}$: unlike other systems such as ferromagnetic $\text{Co}_2\text{MnGa}$ \cite{ChangPRL2017, BelopolskiScience2019} and antiferromagnetic $\text{YMn}_2\text{Ge}_2$ \cite{YangNC2024}, where skyrmions are not an intrinsic bulk property, both topologies are inherent to $\text{Eu}(\text{Ga}_{1-x}\text{Al}_x)_4$, promoting an intrinsic coupling without relying on complex artificial heterostructures. Indeed, our direct observation of magnetic band folding provides evidence that the magnetism (hosting real-space topology) and the electronic structure (hosting momentum-space topology) are strongly coupled. The interplay between skyrmions and $\text{TSSs}$ is predicted to enable conceptually new phenomena, such as the formation of charged skyrmions due to the confinement of $\text{TSSs}$ at the skyrmion radius \cite{HurstPRB2015, AndrikopoulosJAP2016}. This could realize a qualitative change from the conventional current-driven manipulation of skyrmions to electric-field-driven dissipationless control \cite{HurstPRB2015, AndrikopoulosJAP2016}. Moreover, the present results demonstrate that the electronic and magnetic coupling is highly sensitive to the surface termination, appearing pronouncedly only on the $\text{Eu}$-terminated surface. This finding makes $\text{Eu}(\text{Ga}_{1-x}\text{Al}_x)_4$ a platform for controlling the interplay between the two topological facets by designing the surface environment.

In conclusion, we reported a surface-sensitive VUV-ARPES study of $\text{EuGa}_2\text{Al}_2 $ and $\text{EuAl}_4$. We uncovered TSSs associated with bulk DNL and their exceptional robustness against a variety of surface and magnetic perturbations. We also found magnetic band folding, which is pronounced on the Eu-terminated surface, providing evidence for a strong and tunable coupling between the magnetism and the electronic structure. These findings firmly establish the $\text{Eu}(\text{Ga}_{1-x}\text{Al}_x)_4$ family as an intrinsically coupled dual-topology system, offering a rare platform to investigate the interplay between two distinct topological realms with high tunability, all in a single material.

\begin{acknowledgments}
This work was supported by JST-CREST (no. JPMJCR18T1), Grant-in-Aid for Scientific Research (JSPS KAKENHI Grant Number JP21H04435, JP23H01115), KEK-PF (Proposal No. 2024S2-001, 2022G652, 2024G141, 2024G136). \add{We thank the University of the Ryukyus Research Facility Center for the use of the helium liquefier for experiments at OIST.} Y.A. and A.H. acknowledge support from GP-Spin at Tohoku University. T.K. and A.H. thank JSPS.\end{acknowledgments}

\newpage

\begin{figure*}[htbp]
\includegraphics[width=165mm]{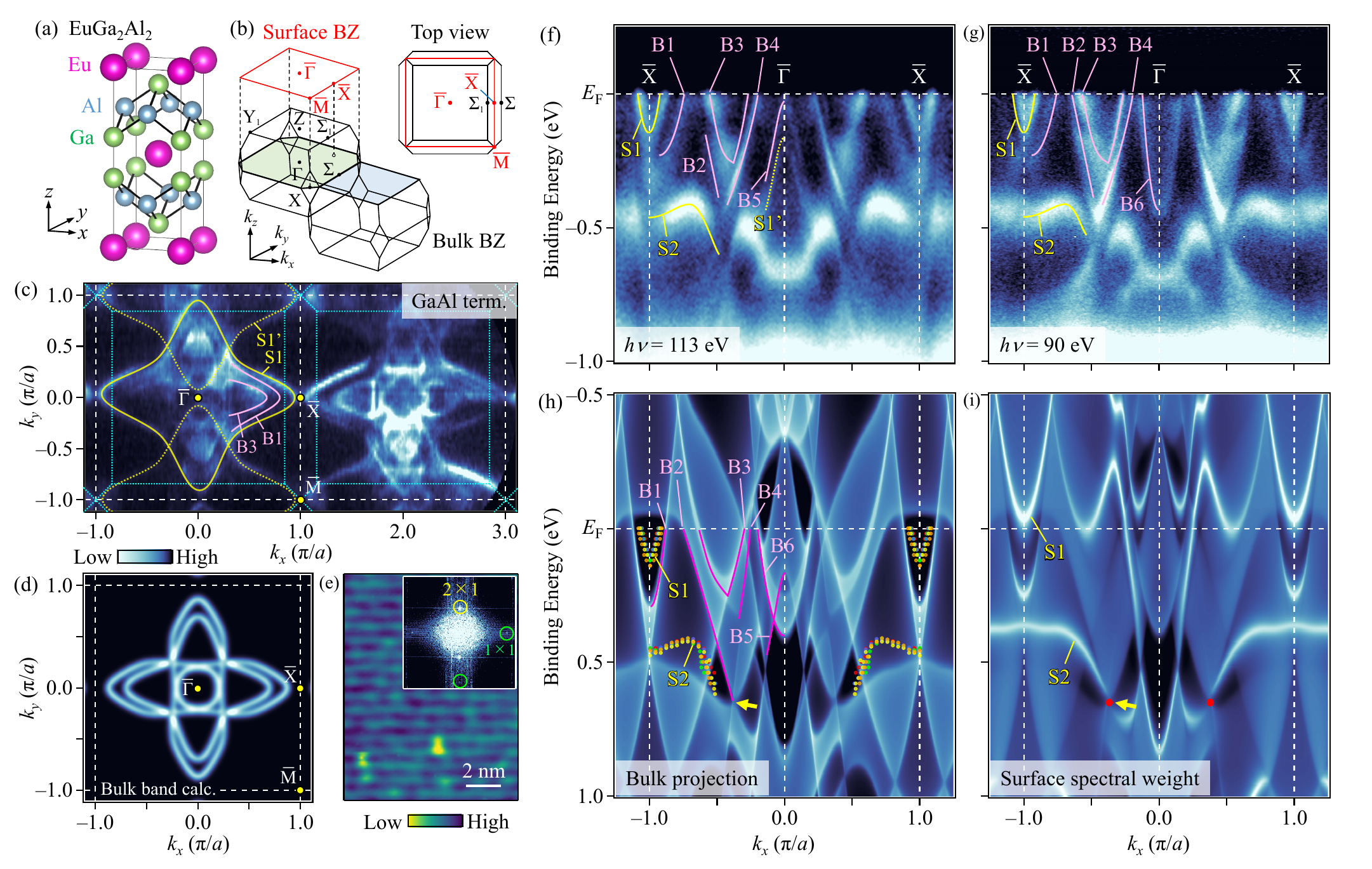}
 \caption{\label{Fig1}
(a) Crystal structure of $\text{Eu}\text{Ga}_2\text{Al}_2$.
(b) Bulk body-centered tetragonal BZ (black) and corresponding surface BZ projected onto the $k_z$ direction (red).
(c) ARPES intensity at $E_{\rm F}$ of the GaAl termination of $\text{Eu}\text{Ga}_2\text{Al}_2$ plotted as a function of $k_x$ and $k_y$ measured at $T$ = 30 K with 113-eV photons. The white dashed lines indicate the surface BZ.
(d) Theoretical calculation of the bulk FS at $k_z = \pi$ plane (Z-$\Sigma_1$-Y$_1$ plane).
(e) STM topographic image of the GaAl termination of $\text{Eu}\text{Ga}_2\text{Al}_2$. The inset shows the corresponding Fast-Fourier-transform patterns, confirming the $2\times1$ superstructure. STM setup condition ($V$/$I$): $-$50 mV/500 pA.
(f) and (g) ARPES intensity along the $k_x$ axis measured at $h\nu$ = 113 eV and 90 eV (which correspond to $k_z \sim 0$ and $\pi$, respectively, according to the inner potential value of 16 eV\add{; see Section 4 in Supplemental Material \cite{SM}}). Solid lines are a guide for the eyes to trace the experimental band dispersions.
(h) Calculated bulk bands projected onto the (001) surface. Spin-orbit coupling was neglected. Circles are the peak position of ARPES spectra obtained at various $h\nu$'s (see Section \add{6} in Supplemental Material for details \cite{SM}).
(i) Surface spectrum $A(k, E_{\rm B})$. The red dot indicates bulk-band touching point where the S2 band is terminated.
}
\end{figure*}

\newpage

\begin{figure}[htbp]
\includegraphics[width=90mm]{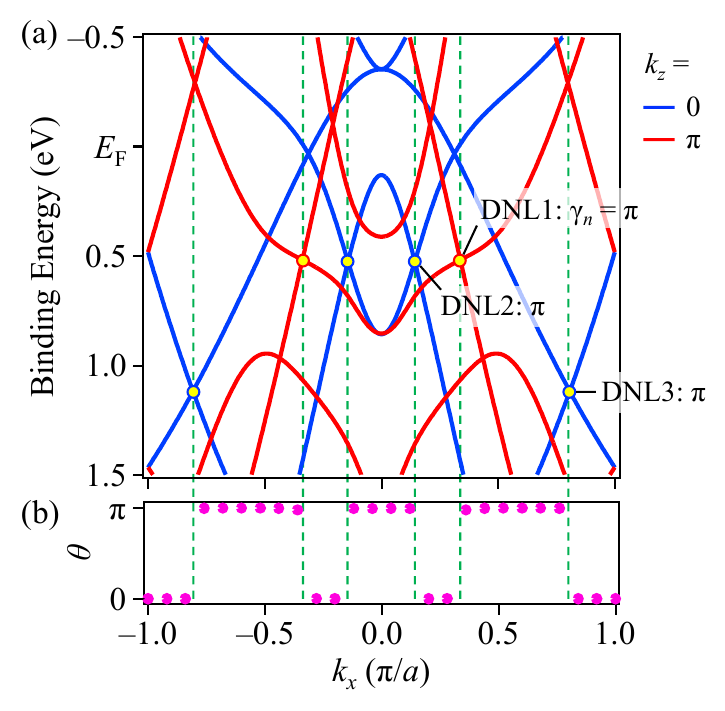}
 \caption{\label{Fig2}
\add{(a) Calculated bulk band dispersions at $k_z$ = 0 (blue) and $\pi$ (red) for EuGa$_2$Al$_2$, together with the Berry phase $\gamma_n$ calculated for paths encircling Dirac nodal lines (DNL1--DNL3). (b) Calculated Zak phase $\theta$ plotted as a function of $k_x$.}}
\end{figure}

\newpage

\begin{figure}[htbp]
\includegraphics[width=120mm]{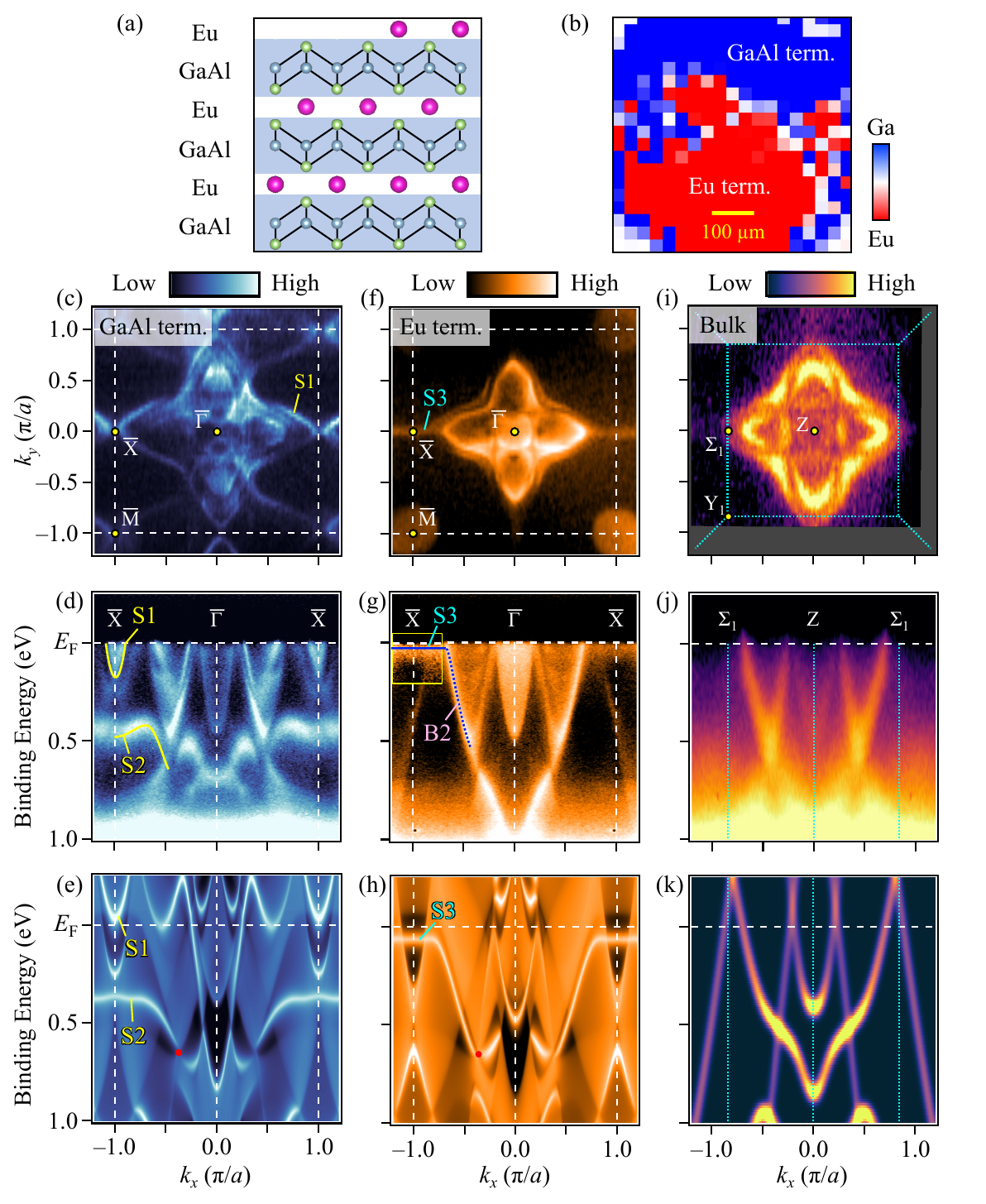}
 \caption{\label{Fig3}
(a) Side view of the crystal, illustrating the two possible surface terminations (Eu monolayer or GaAl tetrahedral layer).
(b) Spatial map of the photoemission intensity of the Ga-3$d$ core-level peaks measured at $h\nu$ = 113 eV. The yellow scale bar represents 100 $\mathrm{\mu}$m.
(c) FS map, (d) band dispersion along the $\bar{\Gamma}$-$\bar{\rm X}$ cut, and (e) calculated surface spectral weight for the GaAl terminated surface of $\text{Eu}\text{Ga}_2\text{Al}_2$.
(f)--(h) Same as (c)--(e) but for the Eu-terminated surface. \add{The $k_z$ value was set to $\pi$.} Image contrast in the $(E, k)$ region enclosed by yellow rectangle was enhanced to make the S3 band clearer.
(i)--(k) Same as (c)--(e) but measured at $h\nu$ = 442 eV\add{, which probes $k_z \sim \pi$}.}
\end{figure}

\newpage

\begin{figure}[htbp]
\includegraphics[width=120mm]{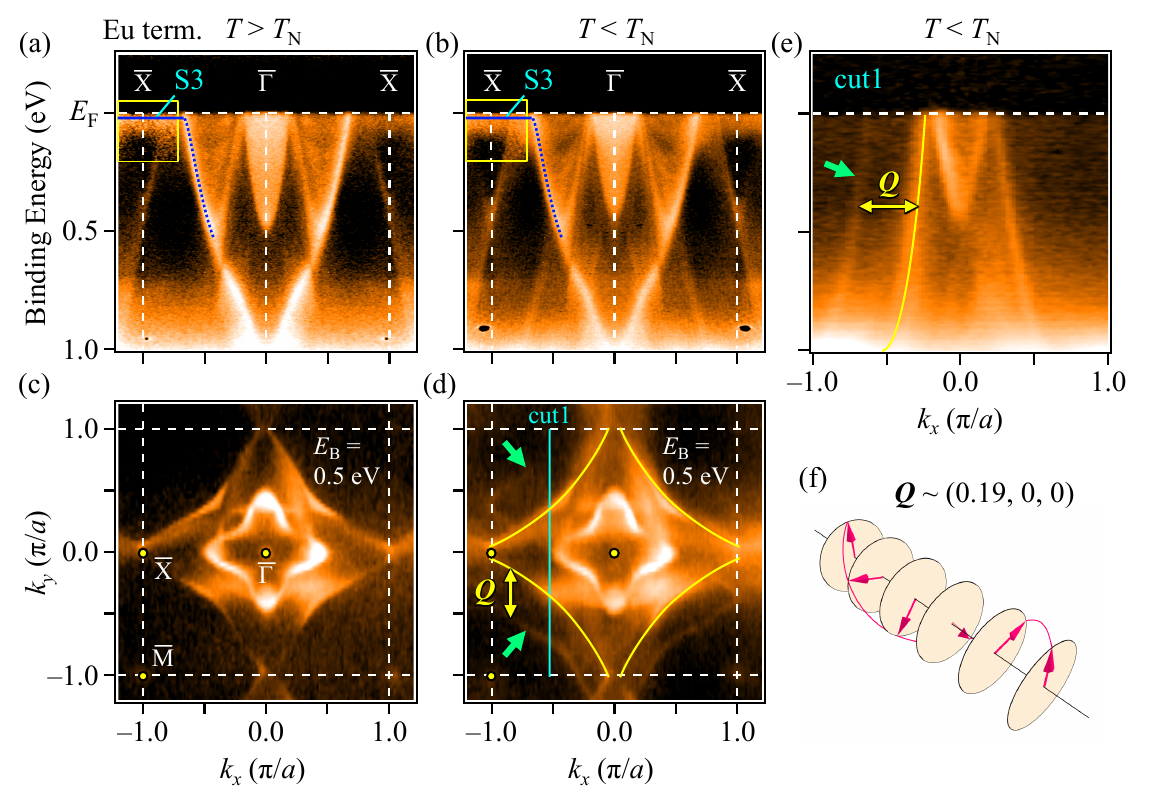}
 \caption{\label{Fig4}
(a) and (b) ARPES-intensity plots measured at $T$ = 30 K ($> T_{\rm N}$) and 7 K ($< T_{\rm N}$), respectively, with the incident energy $h\nu$ = 90 eV along the $\bar{\Gamma}$-$\bar{\rm X}$ high-symmetry path.
(c) and (d) Contour plots at $E_{\rm B}$ = 0.5 eV measured at $T$ = 30 K and 7 K, respectively. The solid yellow lines are a guide for the eyes to trace the original band. Replica bands are marked by green arrows.
(e) ARPES intensity measured along the cut1 (light blue line) in (d).
(f) Schematic of helical magnetic order \cite{MoyaPRM2022,VibhakarPRB2023}.}
\end{figure}

\newpage

\begin{figure}[htbp]
\includegraphics[width=120mm]{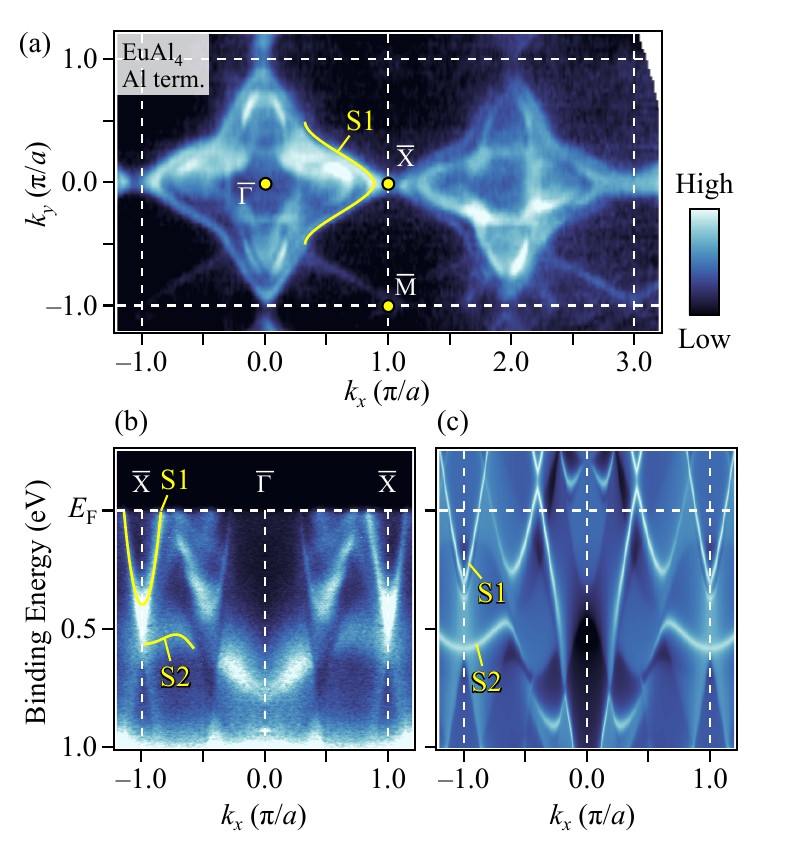}
 \caption{\label{Fig5}
(a) ARPES-intensity mapping at $E_{\rm F}$ of the Al-terminated $\text{EuAl}_4$ surface plotted as a function of the in-plane wave vectors $k_x$ and $k_y$ measured at $h\nu$ = 110 eV.
(b) ARPES-intensity plot measured along the $\bar{\Gamma}$-$\bar{\rm X}$ cut.
(c) Calculated surface spectral function.}
\end{figure}


\section*{References}
\begin{thebibliography}{99}
\bibitem{RoblerNat2006} U. K. R\"{o}ssler, A. N. Bogdanov, and C. Pfleiderer, Nature \textbf{442}, 797 (2006).
\bibitem{MuhlbauerScience2009} S. M\"{u}hlbauer, B. Binz, F. Jonietz, C. Pfleiderer, A. Rosch, A. Neubauer, R. Georgii, and P. B\"{o}ni, Science \textbf{323}, 1166767 (2009).
\bibitem{NagaosaNatTech2013} N. Nagaosa and Y. Tokura, Nat. Nanotech. \textbf{8}, 899 (2013).
\bibitem{LeePRL2009} M. Lee, N. Kang, Y. Onose, Y. Tokura, and N. P. Ong, Phys. Rev. Lett. \textbf{102}, 186601 (2009).
\bibitem{NeubauerPRL2009} A. Neubauer, C. Pfleiderer, B. Binz, A. Rosch, R. Ritz, P. G. Niklowitz, and P. B\"{o}ni, Phys. Rev. Lett. \textbf{102}, 186602 (2009).
\bibitem{ShiomiPRB2009} Y. Shiomi, N. Kanazawa, K. Shibata, Y. Onose, and Y. Tokura, Phys. Rev. B \textbf{88}, 064409 (2013).
\bibitem{FertNatTech2013} A. Fert, V. Cros, and J. Sampaio, Nat. Nanotech. \textbf{8}, 152 (2013).
\bibitem{SampaioNaTech2013} J. Sampaio, V. Cros, S. Rohart, A. Thiaville, and A. Fert, Nat. Nanotech. \textbf{8}, 839 (2013).
\bibitem{Gobel2021Beyond} B. G\"{o}bel, I. Mertig, and O. A. Tretiakov, Physics Reports \textbf{895}, 1 (2021).
\bibitem{KanePRL1} C. L. Kane and E. J. Mele, Phys. Rev. Lett. \textbf{95}, 146802 (2005).
\bibitem{KanePRL2} C. L. Kane and E. J. Mele, Phys. Rev. Lett. \textbf{95}, 226801 (2005).
\bibitem{BernevigPRL2006} B. A. Bernevig and S.-C. Zhang, Phys. Rev. Lett. \textbf{96}, 106802 (2006).
\bibitem{XiaNP2009} Y. Xia, D. Qian, D. Hsieh, L. Wray, A. Pal, H. Lin, A. Bansil, D. Grauer, Y. S. Hor, R. J. Cava, and M. Z. Hasan, Nat. Phys. \textbf{5}, 398 (2009).
\bibitem{ChenScience2009} Y. L. Chen, J. G. Analytis, J.-H. Chu, Z. K. Liu, S.-K. Mo, X. L. Qi, H. J. Zhang, D. H. Lu, X. Dai, Z. Fang, S. C. Zhang, I. R. Fisher, Z. Hussain, and Z.-X. Shen, Science \textbf{325}, 178 (2009).
\bibitem{HasanRMP2010} M. Z. Hasan and C. L. Kane, Rev. Mod. Phys. \textbf{82}, 3045 (2010).
\bibitem{QiRMP2011} X.-L. Qi and S.-C. Zhang, Rev. Mod. Phys. \textbf{83}, 1057 (2011).
\bibitem{AndoJPSJ2013} Y. Ando, J. Phys. Soc. Jpn. \textbf{82}, 102001 (2013).
\bibitem{ArmitageRMP2018} N. P. Armitage, E. J. Mele, and Ashvin Vishwanath, Rev. Mod. Phys. \textbf{90}, 015001 (2018).
\bibitem{Sato2017Review} M. Sato and Y. Ando, Rep. Prog. Phys. \textbf{80}, 076501 (2017).
\bibitem{HurstPRB2015} H. M. Hurst, D. K. Efimkin, J. Zang, and V. Galitski, Phys. Rev. B \textbf{91}, 060401(R) (2015).
\bibitem{AndrikopoulosJAP2016} D. Andrikopoulos, B. Sore, and J. De Boeck, J. Appl. Phys. \textbf{119}, 193903 (2016).
\bibitem{KurebayashiPRR} D. Kurebayashi and O. A. Tretiakov, Phys. Rev. Research \textbf{4}, 043105 (2022).
\bibitem{WuAdvMat2020} H. Wu, F. Gro\ss, B. Dai, D. Lujan, S. A. Razavi, P. Zhang, Y. Liu, K. Sobotkiewich, J. Förster, M. Weigand, G. Sch\"{u}tz, X. Li, J. Gr\"{a}fe, and K. L. Wang, Adv. Funct. Mater. \textbf{32}, 2003380 (2020).
\bibitem{MoyaPRB2023} J. M. Moya, J. Huang, S. Lei, K. Allen, Y. Gao, Y. Sun, M. Yi, and E. Morosan, Phys. Rev. B \textbf{108}, 064436 (2023).
\bibitem{NakamuraJPSJ2015} A. Nakamura, T. Uejo, F. Honda, T. Takeuchi, H. Harima, E. Yamamoto, Y. Haga, K. Matsubayashi, Y. Uwatoko, M. Hedo, T. Nakama, and Y. Onuki, J. Phys. Soc. Jpn. \textbf{84}, 124711 (2015).
\bibitem{TakagiNC2022} R. Takagi, N. Matsuyama, V. Ukleev, L. Yu, J. S. White, S. Francoual, J. R. L. Mardegan, S. Hayami, H. Saito, K. Kaneko, K. Ohishi, Y. \=Onuki, T. Arima, Y. Tokura, T. Nakajima, and S. Seki, Nat. Commun. \textbf{13}, 1472 (2022).
\bibitem{LeiNC2023} S. Lei, K. Allen, J. Huang, J. M. Moya, T. C. Wu, B. Casas, Y. Zhang, J. S. Oh, M. Hashimoto, D. Lu, J. Denlinger, C. Jozwiak, A. Bostwick, E. Rotenberg, L. Balicas, R. Birgeneau, M. S. Foster, M. Yi, Y. Sun, and E. Morosan, Nat. Commun. \textbf{14}, 5812 (2023).
\bibitem{ShangPRB2021} T. Shang, Y. Xu, D. J. Gawryluk, J. Z. Ma, T. Shiroka, M. Shi, and E. Pomjakushina, Phys. Rev. B \textbf{103}, L020405 (2021).
\bibitem{ShangReview2024} T. Shang, Y. Xu, S. Gao, R. Yang, T. Shiroka, and M. Shi, J. Phys.: Condens. Matter. \textbf{37}, 013002 (2024).
\bibitem{AraiNC2026} Y. Arai, K. Nakayama, A. Honma, S. Souma, D. Shiga, H. Kumigashira, T. Takahashi, K. Segawa, and T. Sato, Nat. Commun. \textbf{17}, 3162 (2026).
\bibitem{HayamiJPSJ2022} S. Hayami, J. Phys. Soc. Jpn. \textbf{91}, 023705 (2022).
\bibitem{HayamiJPM2023} S. Hayami, J. Phys. Mater. \textbf{6}, 014006 (2023).
\bibitem{KobataJPSJ2016} M. Kobata,, S.-i. Fujimori, Y. Takeda, T. Okane, Y. Saitoh, K. Kobayashi, H. Yamagami, A. Nakamura, M. Hedo, T. Nakama, and Y. Onuki, J. Phys. Soc. Jpn. \textbf{85}, 094703 (2016).
\bibitem{EatonPRB2024} A.Eaton, B. Kuthanazhi, P. C. Canfield, B. Schrunk, N. H. Jo, Y. Kushnirenko, E. O'Leary, L.-L. Wang, and A. Kaminski, Phys. Rev. B \textbf{110}, 125150 (2024).
\bibitem{MiaoPRX2024} H. Miao, J. Bouaziz, G. Fabbris, W. R. Meier, F. Z. Yang, H. X. Li, C. Nelson, E. Vescovo, S. Zhang, A. D. Christianson, H. N. Lee, Y. Zhang, C. D. Batista, and S. Bl\"{u}gel, Phys. Rev. X \textbf{14}, 011053 (2024).
\bibitem{LiArXiv2025} T. Li, L. Chen, J. Yuan, Z. Liu, Y. Yang, Z. Jiang, J. Ding, J. Liu, J. Liu, Z. Sun, Y. Guo, T. Zhang, and D. Shen, preprint at https://arxiv.org/abs/2509.04742.
\bibitem{SM} See Supplemental Material at (URL), which includes Refs. \cite{StavinohaPRB2018, AraiNC2026, KitamuraRSI2022, Giannozzi2009, Giannozzi2014, Corso2014, Wannier90, WannierTools, VESTA}, for details of the experimental conditions, \add{mesh grid noise reduction,} band calculations, photon energy dependence of ARPES spectra, \add{calculated FS, roles of SOC and magnetism,} core level spectra, additional STM data, and magnetic band folding.
\bibitem{StavinohaPRB2018} M. Stavinoha, J. A. Cooley, S. G. Minasian, T. M. McQueen, S. M. Kauzlarich, C.-L. Huang, and E. Morosan, Phys. Rev. B \textbf{97}, 195146 (2018).
\bibitem{MoyaPRM2022} J. M. Moya, S. Lei, E. M. Clements, C. S. Kengle, S. Sun, K. Allen, Q. Li, Y. Y. Peng, and A. A. Husain, Phys. Rev. Mater. \textbf{6}, 074201 (2022).
\bibitem{VibhakarPRB2023} A. M. Vibhakar, D. D. Khalyavin, J. M. Moya, P. Manuel, F. Orlandi, S. Lei, E. Morosan, and A. Bombardi, Phys. Rev. B \textbf{108}, L100404 (2023).
\bibitem{SchoopNC2016} L. M. Schoop, M. N. Ali, C. Stra\ss er, A. Topp, A. Varykhalov, D. Marchenko, V. Duppel, S. S. P. Parkin, B. V. Lotsch, and C. R. Ast, Nat. Commun. \textbf{7}, 11696 (2016).
\bibitem{TakanePRB2016} D. Takane, Z. Wang, S. Souma, K. Nakayama, C. X. Trang, T. Sato, T. Takahashi, and Y. Ando, Phys. Rev. B \textbf{94}, 121108(R) (2016).
\bibitem{BoukhvalovAFM2019} D. W. Boukhvalov, R. Edla, A. Cupolillo, V. Fabio, R. Sankar, Y. Zhu, Z. Mao, J. Hu, P. Torelli, G. Chiarello, L. Ottaviano, and A. Politano, Adv. Funct. Mater. \textbf{29}, 1900438 (2019).
\bibitem{ChangPRL2017} G. Chang, S.-Y. Xu, X. Zhou, S.-M. Huang, B. Singh, B. Wang, I. Belopolski, J. Yin, S. Zhang, A. Bansil, H. Lin, and M. Z. Hasan, Phys. Rev. Lett. \textbf{119}, 156401 (2017).
\bibitem{BelopolskiScience2019} I. Belopolski, K. Manna, D. S. Sanchez, G. Chang, B. Ernst, J. Yin, S. S. Zhang, T. Cochran, N. Shumiya, H. Zheng, B. Singh, G. Bian, D. Multer, M. Litskevich, X. Zhou, S.-M. Huang, B. Wang, T.-R. Chang, S.-Y. Xu, A. Bansil, C. Felser, H. Lin, and M. Z. Hasan, Science \textbf{365}, 1278 (2019).
\bibitem{YangNC2024} X. P. Yang, Y.-T. Yao, P. Zheng, S. Guan, H. Zhou, T. A. Cochran, C.-M. Lin, J.-X. Yin, X. Zhou, Z.-J. Cheng, Z. Li, T. Shi, M. S. Hossain, S. Chi, I. Belopolski, Y.-X. Jiang, M. Litskevich, G. Xu, Z. Tian, A. Bansil, Z. Yin, S. Jia, T.-R. Chang, and M. Z. Hasan, Nat. Commun. \textbf{15}, 7052 (2024).
\bibitem{KitamuraRSI2022} M. Kitamura, S. Souma, A. Honma, D. Wakabayashi, H. Tanaka, A. Toyoshima, K. Amemiya, T. Kawakami, K. Sugawara, K. Nakayama, K. Yoshimatsu, H. Kumigashira, T. Sato, and K. Horiba, Rev. Sci. Instrum. \textbf{93}, 033906 (2022).
\bibitem{Giannozzi2009} P. Giannozzi, S. Baroni, N. Bonini, M. Calandra, R. Car, C. Cavazzoni, D. Ceresoli, G. L. Chiarotti, M. Cococcioni, I. Dabo, A. D. Corso, S. Gironcoli, S. Fabris, G. Fratesi, R. Gebauer, U. Gerstmann, C. Gougoussis, A. Kokalj, M. Lazzeri, L. Martin-Samos, N. Marzari, F. Mauri, R. Mazzarello, S. Paolini, A. Pasquarello, L. Paulatto, C. Sbraccia, S. Scandolo, G. Sclauzero, A. P. Seitsonen, A. Smogunov, P. Umari, and R. M. Wentzcovitch, J. Phys. Condens. Matter. \textbf{21}, 395502 (2009).
\bibitem{Giannozzi2014} P. Giannozzi, O. Andreussi, T. Brumme, O. Bunau, M. B. Nardelli, M. Calandra, R. Car, C. Cavazzoni, D. Ceresoli, M. Cococcioni, N. Colonna, I. Carnimeo, A. D. Corso, S. Gironcoli, P. Delugas, R. A. DiStasio Jr, A. Ferretti, A. Floris, G. Fratesi, G. Fugallo, R. Gebauer, U. Gerstmann, F. Giustino, T. Gorni, J. Jia, M. Kawamura, H.-Y. Ko, A. Kokalj, E. K\"{u}ç\"{u}kbenli, M. Lazzeri, M. Marsili, N. Marzari, F. Mauri, N. L. Nguyen, H.-V. Nguyen, A. Otero-de-la-Roza, L. Paulatto, S. Ponc\'{e}, D. Rocca, R. Sabatini, B. Santra, M. Schlipf, A. P. Seitsonen, A. Smogunov, I. Timrov, T. Thonhauser, P. Umari, N. Vast, X. Wu, and S. Baroni, J. Phys. Condens. Matter. \textbf{29}, 465901 (2017).
\bibitem{Corso2014} A. Dal Corso, Comput. Mater. Sci. \textbf{95}, 337 (2014).
\bibitem{Wannier90} A. A. Mostofi, J. R. Yates, G. Pizzi, Y.-S. Lee, I. Souza, D. Vanderbilt, and N. Marzari, Comput. Phys. Commun. \textbf{185}, 2309 (2014).
\bibitem{WannierTools} Q. S. Wu, S. N. Zhang, H. F. Song, M. Troyer, and A. A. Soluyanov, Comput. Phys. Commun. \textbf{224}, 405 (2018).
\bibitem{VESTA} K. Momma and F. Izumi, J. Appl. Crystallogr. \textbf{44}, 1272 (2011).
\end{thebibliography}
\end{document}